\documentclass[amsmath, amssymb, aps, prx, twocolumn, superscriptaddress, longbibliography]{revtex4-2}

\usepackage[separate-uncertainty, qualifier-mode=text,parse-numbers=false]{siunitx}
\usepackage{graphicx}
\usepackage{braket}
\usepackage[T1]{fontenc}
\usepackage{lmodern}
\usepackage[utf8]{inputenc}
\usepackage{hyperref}
\usepackage{orcidlink}
\usepackage{booktabs}
\usepackage{bm}
\usepackage[normalem]{ulem}
\makeatletter
\g@addto@macro\bfseries{\boldmath}
\makeatother

\hypersetup{
	pdfpagemode=UseOutlines,
	bookmarksnumbered=true,
	pdflang=en,
	hidelinks,
	hypertexnames=false,
}

\DeclareMathOperator{\erf}{erf}
\DeclareSIUnit\gauss{G}
\DeclareSIUnit\OhmSq{\ensuremath{\Omega / \square}}

\begin{document}

\title{Long-lived giant circular Rydberg atoms at room temperature}

\author{Einius Pultinevicius\orcidlink{0009-0005-7404-9178}}
\author{Aaron G\"{o}tzelmann\orcidlink{0000-0001-5527-5878}}
\author{Fabian Thielemann\orcidlink{0000-0003-4578-8500}}
\author{Christian H\"{o}lzl\orcidlink{0000-0002-2176-1031}}
\author{Florian Meinert\orcidlink{0000-0002-9106-3001}}
\altaffiliation{Corresponding author: \href{mailto:f.meinert@physik.uni-stuttgart.de}{f.meinert@physik.uni-stuttgart.de}}

\affiliation{5. Physikalisches Institut and Center for Integrated Quantum Science and Technology, Universit\"{a}t Stuttgart, Pfaffenwaldring 57, 70569 Stuttgart, Germany}


\begin{abstract}
	Stability achieved by large angular momentum is ubiquitous in nature, with examples ranging from classical mechanics, over optics and chemistry, to nuclear physics.
	In atoms, angular momentum can protect excited electronic orbitals from decay due to selection rules.
	This manifests spectacularly in highly excited Rydberg states.
	Low angular momentum Rydberg states are at the heart of recent breakthroughs in quantum computing, simulation and sensing with neutral atoms.
	For these applications the lifetime of the Rydberg levels sets fundamental limits for gate fidelities, coherence times, or spectroscopic precision.
	The quest for longer Rydberg state lifetimes has motivated the generation, coherent control and trapping of circular Rydberg atoms, which are characterized by the maximally allowed electron orbital momentum and were key to seminal experiments with single atoms and photons.
	Here, we report the observation of individually trapped circular Rydberg atoms with lifetimes of more than 10 milliseconds, two orders of magnitude longer-lived than the established low angular momentum orbitals.
	This is achieved via Purcell suppression of blackbody modes at room temperature.
	We coherently control individual circular Rydberg levels at so far elusive principal quantum numbers of up to $n=103$, and observe tweezer trapping of the Rydberg atoms on the hundred millisecond scale.
	Our results pave the way for quantum information processing and sensing utilizing the combination of extreme lifetimes and giant Rydberg blockade.
	\end{abstract}
\maketitle
\newpage

\section*{Introduction}\label{Introduction}

Systems of neutral atoms trapped in optical tweezer arrays evolved over the past few years into one of the most promising platforms for quantum computing and quantum simulation \cite{Browaeys2020,Adams2020}. The unprecedented control over interactions in these arrays, mediated by laser excitation to highly excited Rydberg states, made it possible to encode and simulate quantum many-body Hamiltonians of paradigmatic lattice models \cite{Leseleuc2019,Semeghini2021,Scholl2021,Ebadi2021,Qiao2025}. In the context of digital quantum computing, Rydberg states are a key ingredient to perform high-fidelity gates between two or more atoms \cite{Evered2023,Tsai2025} to operate large-scale quantum processors \cite{Bluvstein2024,Graham2022}. A fundamental limitation for coherence times or gate fidelities, which inevitably comes with the commonly used Rydberg states ($S$ or $D$ orbitals), is their finite lifetime on the order of \SI{100}{\micro\second} due to optical and blackbody radiation (BBR) induced decay \cite{Saffman2010}. Circular Rydberg states (CRS) \cite{Hulet1983}, characterized by a maximal angular momentum (i.e., $|m|=l=n-1$, where $l$ and $m$ are the orbital and magnetic quantum numbers, and $n$ is the principal quantum number), provide means to overcome this limitation \cite{Nguyen2018,Meinert2020,Cohen2021}. 
These states exhibit exceptionally long lifetimes due to the strong inhibition of spontaneous optical decay by selection rules~\cite{SobelmanBook}. This inhibition originates from its large angular momentum~\cite{Dunning2009}, a stabilization mechanism that is omnipresent in nature~\cite{Passaro2017, Gregg2015, Knoop2008, JainBook}. 
While optical decay for these highest possible angular momentum states is inhibited by selection rules \cite{SobelmanBook}, BBR also needs to be suppressed, e.g., via a cryogenic environment as in the Nobel prize-winning works on single-atom single-photon interaction \cite{Haroche2013}.

Here, we observe lifetimes of strontium circular Rydberg atoms which exceed \SI{10}{\milli\second} in a room temperature neutral atom tweezer experiment. Instead of cryogenic cooling~\cite{CantatMoltrecht2020, Zhang2025, Jin2026}, we modify the mode density at microwave frequencies at the position of the atom \cite{Kleppner1981,Vaidyanathan1981,Cohen2021,Nguyen2018,Meinert2020}. This causes a more than 20-fold suppression of BBR-induced decay, and effectively provides lifetimes that in free space would require a \SI{14}{K} environment. Further, we demonstrate coherent excitation of large CRS up to $n=103$ (\SI{1.1}{\micro\meter} orbit diameter) using microwave control over a ladder of Rydberg levels, which allows us to characterize the blackbody suppression and Rydberg state lifetime enhancement across a wide range of frequencies.

Coherent interaction control between circular Rydberg atoms has recently been demonstrated at $n\approx50$ \cite{Mehaignerie2025}. For comparison, our high $n$ brings about a thousand-fold stronger van der Waals blockade, and more than ten times larger dipole-exchange coupling.
Under the condition that the technical factors limiting the currently achieved dephasing time~\cite{Hoelzl2024} are addressed, these results push the fundamental barrier (i.e., the Rydberg state's $T_1$ time) for establishing quantum simulation on the Rydberg platform on timescales two orders of magnitude longer than what is currently possible.
Reduced blackbody decay to other Rydberg states is also beneficial for erasure conversion schemes \cite{Scholl2023,Ma2023}, and mitigates anomalous dephasing in atom arrays \cite{Goldschmidt2016,Festa2022}.

\begin{figure*}[tbh!]
	\centering
	\includegraphics[scale=1]{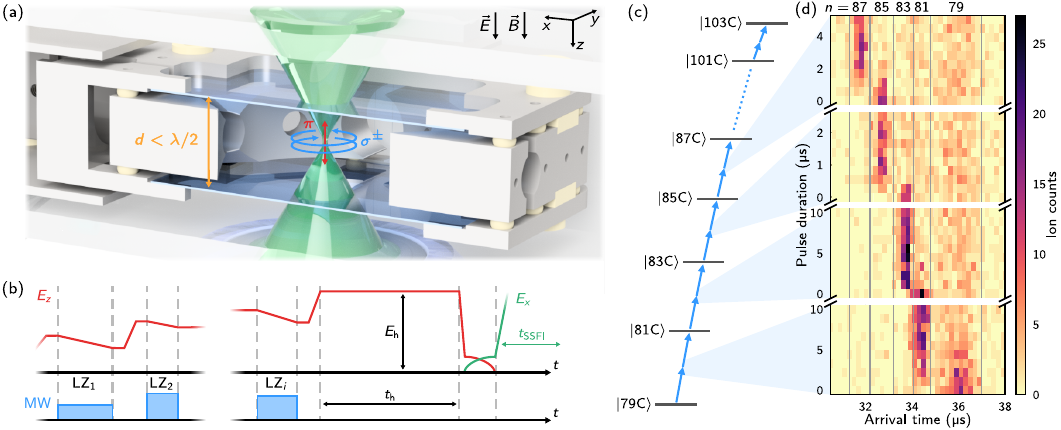}
	\caption{Suppression capacitor and coherent high-$n$ circular state control. (a) Experimental setup:
	A circular Rydberg atom is trapped in an optical tweezer at the center of an electrode structure.
	This structure consists of four ring-shaped electrodes in the $xy$-plane in between two indium tin oxide (ITO)-coated glass plates, the latter forming the capacitor with spacing $d$ for suppression of $\sigma^\pm$ polarized blackbody radiation modes of wavelength $\lambda$.
	Applied electric and magnetic fields along $z$ (quantization axis) orient the circular Rydberg orbit so that it lies in the $xy$-plane. 
	(b) Sketch of the succession of Landau-Zener microwave (MW) pulses ($\mathrm{LZ}_i$) and the electric field sequence ($E_z$) applied for preparing CRS up to $n=103$ starting from $\ket{79\mathrm{C}}$. 
	Each pulse $\mathrm{LZ}_i$ drives a two-photon transition in the simplified level scheme depicted in (c), while $E_z$ is ramped linearly across the resonance condition for coherent Landau-Zener population transfer.
	After a hold time $t_\mathrm{h}$ at a fixed field $E_\mathrm{h}$, the final Rydberg state is detected via state-selective field ionization (SSFI) by applying a large field $E_x$ ramp.
(d) SSFI measurements of Landau-Zener sweeps of four exemplary transitions up to $n=87$, showing adiabatic transfer ($\ket{n\mathrm{C}} \leftrightarrow \ket{n+2\mathrm{C}}$) when the $\mathrm{LZ}_i$ pulse duration is sufficiently long.}
	\label{fig:Fig1}
\end{figure*}

In our experiment, we consider CRS $\ket{n\mathrm{C}}$ trapped in the center between two conducting plates of an in-vacuum capacitor (Fig.~\ref{fig:Fig1}(a)).
The capacitor is formed by a pair of transparent glass plates coated on the inside with indium-tin oxide (ITO) to achieve high reflectivity in the microwave domain while preserving optical access~\cite{Meinert2020}. 
In free space BBR predominantly causes transitions to CRS with adjacent principal quantum number ($\ket{n\mathrm{C}} \rightarrow \ket{n\pm1\mathrm{C}}$).
In this process, the electron changes its orbital angular momentum by one quantum and emits or absorbs a circularly polarized microwave photon.
The central purpose of the capacitor is to suppress the photon mode density at these BBR transitions. 
This can be achieved when the photon wavelength $\lambda>2d$, where $d$ is the plate distance \cite{Vaidyanathan1981,Nguyen2018,Meinert2020}.
Suppression applies for modes polarized parallel to the plates, and thus for the detrimental circular photons.
In our finite size setup, with further electrodes and conductive elements (see Fig.~\ref{fig:Fig1}(a)), deviations from this idealized model occur due to formation of additional resonant modes. Here, we characterize these effects on the BBR environment by going far beyond the onset of Purcell suppression at $ \lambda \approx 2d$.

For a given mode density suppression, which in detail depends on the capacitor design (see Methods~\ref{sec:purcell}), we exploit a second possibility to enhance the CRS lifetime, namely by increasing the principal quantum number $n$. This is because the rate for BBR-induced transitions to neighboring circular Rydberg levels decreases with $\propto n^{-2}$~\cite{Gallagher1994}. Realization of CRS with $n$ significantly larger than $\approx 50$, however, remained elusive because of increasing state degeneracy ($\propto n^2$), and increasing time required for state preparation. Only recently, our group showed trapped CRS with $n=79$ \cite{Hoelzl2024}. Now we gain access to much higher CRS
by combining adiabatic state preparation with up to 12 subsequent microwave pulses. The pulses coherently link CRS from $n=79$ to $103$ (Fig.~\ref{fig:Fig1}(c)).

\section*{Results}
\subsection*{Large circular states with $n>100$}

To this end, our experiments start with the preparation of a single trapped $^{88}$Sr atom inside our capacitor structure (Fig.~\ref{fig:Fig1}(a))~\cite{Hoelzl2023}. 
We promote the atom to the $\ket{79\mathrm{C}}$ CRS by a combination of optical excitation and adiabatic state transfer with a circularly polarized radio-frequency drive applied to the remaining four ring electrodes of our electrode structure (details are described in Ref.~\cite{Hoelzl2024}). 
From there on, the principal quantum number of the circular orbit is successively increased in steps of two by a series of two-photon microwave transitions. 
To achieve robust coherent state transfer up to $\ket{103\mathrm{C}}$, we apply adiabatic Landau-Zener (LZ) sweeps through the two-photon resonance for each of the transitions (Fig.~\ref{fig:Fig1}(d)). 
The Rydberg state is finally read out using state-selective field-ionization (SSFI) (Fig.~\ref{fig:Fig1}(b)), and detection of the produced ions on a microchannel plate detector (MCP). The arrival time at the MCP is then mapped to the principal quantum number of the ionized CRS (Fig.~\ref{fig:Fig1}(d)).

\begin{figure}[tbh!]
	\centering
	\includegraphics[scale=1]{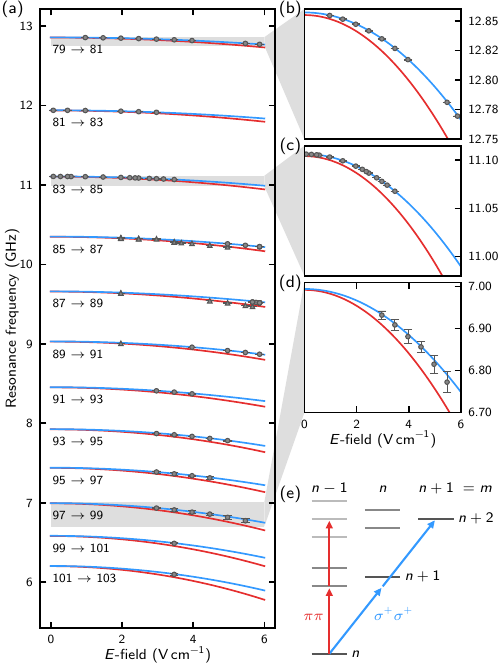}
	\caption{Circular state microwave spectra. (a) Measured microwave resonance frequencies for two-photon transitions as a function of the applied electric field along $z$, with principal quantum number ranging from $n=79$ to $n=103$. 
	Circles (triangles) show data for transitions from $\ket{n\mathrm{C}}$ to circular (elliptical) states with $n+2$. 
	Solid lines show the prediction from a hydrogen model for $\sigma^+\sigma^+$ transitions between CRS (blue), and the nearby $\pi\pi$ transitions connecting a CRS to an elliptical state (red) as indicated in the level scheme shown in (e). The zoom-ins (b)--(d) demonstrate the spectroscopic resolution between both second-order Stark-shifted transitions in the electric field, which allows us to selectively address $\sigma^+\sigma^+$ transitions for high-$n$ CRS preparation. Errorbars show the FWHM of the measured resonance signals.}
	\label{fig:Fig2}
\end{figure}

The established control over many high-$n$ circular Rydberg levels requires extensive microwave spectroscopy to identify and optimize all necessary transitions (Fig.~\ref{fig:Fig2}(a)).
The target transitions ($\ket{n\mathrm{C}} \leftrightarrow \ket{n+2\mathrm{C}}$, blue arrows in Fig.~\ref{fig:Fig2}(e)) need to be isolated from resonances to other non-circular states. 
Most transitions are indeed shifted far out of resonance by their linear Stark and Zeeman shifts in the applied electric and magnetic quantization fields, except for a transition to an elliptical Rydberg level (red arrows in Fig.~\ref{fig:Fig2}(e)).
To separate the two, we use the difference of their quadratic Stark shifts and implement the Landau-Zener sweeps in electric fields of a few \si{\volt\per\centi\meter}.
We choose those field values carefully to avoid complications from much stronger resonant single-photon processes (see Methods~\ref{sec:spectroscopy}). 
The individual Landau-Zener transitions are then implemented with a fixed microwave frequency by ramping the electric field across the two-photon resonance.

\subsection*{Long circular state lifetime via Purcell suppression of BBR}

\begin{figure*}[tb!]
	\centering
	\includegraphics[scale=1]{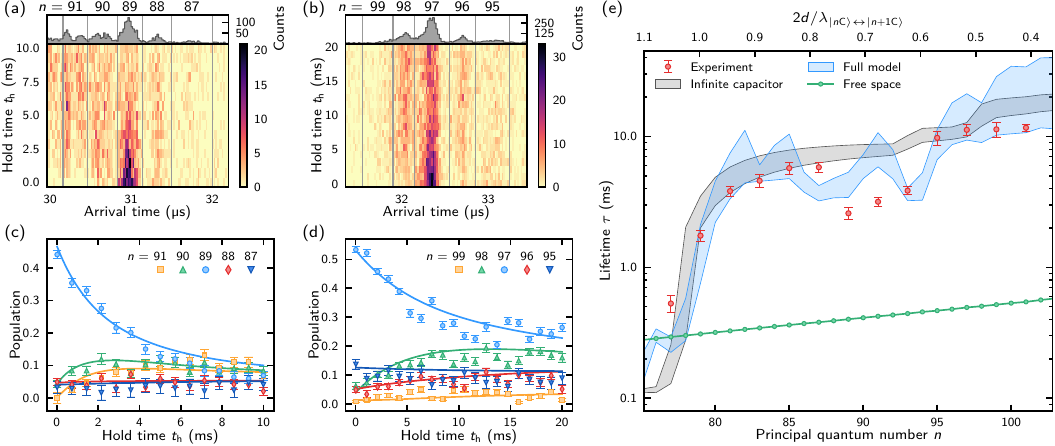}
	\caption{Circular Rydberg state lifetimes. (a, b) SSFI histograms of the arrival time distribution on the MCP as a function of the hold time $t_\mathrm{h}$, for atoms initially prepared in $\ket{89\mathrm{C}}$ (a) and $\ket{97\mathrm{C}}$ (b). 
		Vertical gray lines indicate the bins used to distinguish Rydberg manifolds with principal quantum numbers $n$ as indicated on top.
		The top row shows the histogram integrated over $t_\mathrm{h}$. 
		(c), (d) Relative state populations in the different $n$-manifolds (symbols with $n$ as indicated) evaluated from the histograms in (a) and (b), respectively.
		For better visibility only five $n$-values around the initial state are depicted. Solid lines show a rate model fit to the data, from which the lifetime of the initial CRS is extracted. 
		The data in (a) and (b) is obtained from 4,000 realizations of the experiment for each value of $t_\mathrm{h}$.
		(e) Measured CRS lifetime (red circles) as a function of $n$, obtained from data sets as shown in (c), (d) for different initial states $\ket{n\mathrm{C}}$ (data point for $n=77$ from \cite{Hoelzl2024}). 
		The top abscissa indicates the corresponding criteria $2d/\lambda$ for $\sigma^+$ blackbody suppression, with $\lambda$ the microwave transition wavelength to $\ket{n+1\mathrm{C}}$. 
		Calculated lifetimes in free space (green), in an infinitely extended capacitor with finite reflectivity (gray), and in our full simulated electrode structure (blue) at a temperature of \SI{300}{\kelvin} are shown for comparison. 
		Shaded areas of the capacitor calculations account for uncertainties in the ITO plate distance and reflectivity (see text).
		Errorbars in (c), (d) and (e) depict $1\sigma$-confidence intervals.}
	\label{fig:Fig3}
\end{figure*}

With this control over a wide range of CRS, we are in a position to measure their lifetimes and to investigate the effect of the BBR suppression capacitor. 
Changing the principal quantum number allows us to tune the frequency (\textit{cf.} Fig.~\ref{fig:Fig2}) and wavelength of the circularly polarized blackbody photons which induce the strong transitions to the next lower and higher lying CRS $\ket{n\pm1\mathrm{C}}$ ($\sigma^+$ transitions with $\Delta m =  \pm 1$). 
With the plate distance in our setup, $d=\SI{10.5(2)}{\milli\meter}$, and the approximate criteria for suppression of blackbody photons, $\lambda>2d$, one may expect enhancement of the Rydberg lifetime for $n \gtrsim 78$. 
To measure the CRS lifetime, we introduce a variable hold time $t_\mathrm{h}$ between state preparation and Rydberg state readout via SSFI and ion detection (\textit{cf.} Fig~\ref{fig:Fig1}(b)). During this time, an applied electric field $E_\mathrm{h} = \SI{5.9}{\volt\per\centi\meter}$ ensures that the circular orbit is oriented parallel to the capacitor plates.
Fig.~\ref{fig:Fig3}(a) and (b) reveal how the initial CRS population spreads to neighboring $n$ manifolds as a function of $t_\mathrm{h}$, at the examples of $\ket{89\mathrm{C}}$ and $\ket{97\mathrm{C}}$, respectively. Notably, we can immediately identify lifetimes on the \SI{10}{\milli\second} scale for $\ket{97\mathrm{C}}$ and significantly faster decay for $\ket{89\mathrm{C}}$.

We plot the time dependent population in the different $n$-manifolds in Fig.~\ref{fig:Fig3}(c) and (d) respectively.
To extract the CRS lifetime, a bare exponential fit to the population of the initial state is not sufficient.
This is because the population dynamics is affected by secondary transitions back into the initial $n$-manifold \cite{Hoelzl2024, Wu2023, CantatMoltrecht2020}. 
We take this into account by describing the decay dynamics between the $n$-manifolds with a rate equation model (see Methods~\ref{sec:rate_model}).
Fitting this rate model to datasets as shown in Fig.~\ref{fig:Fig3}(c) and (d) taken at different values of $n$, allows for extracting the CRS lifetime for different initial states $\ket{n\mathrm{C}}$ (Fig.~\ref{fig:Fig3}(e)).
With increasing $n$ we observe first a strong enhancement effect by the capacitor on the lifetimes around $n \approx 80$. 
This stems from the suppression of $\sigma^+$ photons inducing transitions to $\ket{n\pm1\mathrm{C}}$ as discussed above, yielding lifetimes $\approx \SI{5}{\milli\second}$ in the region around $n=85$.

When these strongest transitions become suppressed, the about 160 ($n=80$) to 200 ($n=100$) times  weaker coupling to elliptical states of the next higher $n$ contributes significantly to the CRS lifetime (see Methods~\ref{sec:rate_model}).
Those are $\pi$ transitions ($\Delta m=0$), which are not suppressed by the boundary conditions of the plate capacitor. 
On the contrary, these modes are even enhanced by our ring electrode configuration for certain frequencies, leading to a dip in the measured lifetime around $n\approx90$ (see Methods~\ref{sec:purcell}).
Notably, the effect of these modes is also clearly visible in the $\ket{89\mathrm{C}}$ data of Fig.~\ref{fig:Fig3}(a), which shows a pronounced asymmetric decay toward higher $n$ manifolds as compared to the data for $\ket{97\mathrm{C}}$.
Additional evidence for the enhanced decay to elliptical states can be seen in the integrated SSFI histogram. These states ionize at slightly different fields compared to CRS, causing a substructure in the $n=91$ and $n=90$ bins (see Methods~\ref{sec:SSFI}).
Finally, increasing $n$ even further to values above $n=95$, we achieve lifetimes of more than \SI{10}{\milli\second}, exceeding the expected lifetime in free space by more than a factor of 20. For example, we measure a lifetime of \SI{11.5(8)}{\milli\second} for $\ket{101\mathrm{C}}$, which is around 21 times larger than the free-space lifetime of \SI{545}{\micro\second}.
Note that a significant effect here is another step in the lifetime enhancement for $n \gtrsim 95$, attributed to the suppression of weak $\sigma^+$ transitions to elliptical states in the  $n+2$ manifold, which are driven by photons with about half the wavelength compared to transitions to $n+1$ (see Methods~\ref{sec:rate_model}).
Additionally, the above-mentioned $\pi$ modes between the ring electrodes are suppressed for the same range of $n$.
The observed lifetimes exceed previous room-temperature results at $n=60$ by an order of magnitude \cite{Wu2023}, and also outperform lifetimes obtained in a cryogenic environment by a factor of three \cite{CantatMoltrecht2020}.

Our experimental findings are supported by detailed numerical analysis of the capacitor structure (see Methods~\ref{sec:purcell}). The trend of the data, including the lifetime enhancement steps around $n \approx 80$ and $n \gtrsim 95$, is relatively well explained by a simplified model assuming an infinitely extended plate capacitor with finite reflectivity (gray shaded region in Fig.~\ref{fig:Fig3}(e)). The decreased lifetime around $n\approx90$, however, is not captured. For a more sophisticated simulation, we model the BBR suppression induced by the full electrode structure (\textit{cf.} Fig.~\ref{fig:Fig1}(a)) using finite-element methods. Effectively, together with the ring-shaped electrodes, the capacitor is now extended to a cylindrical cavity. Such a structure supports additional discrete $\pi$-polarized modes, where the one with the lowest energy and  longest wavelength causes the lifetime reduction around $n\approx90$ (blue shaded region in Fig.~\ref{fig:Fig3}(e)). For the simulations, we use the microwave reflectivity of the ITO plates, $R=\SI{96 \left(^{+1}_{-2}\right)}{\percent}$, and the plate distance $d=\SI{10.5(2)}{\milli\meter}$. The shaded regions account for the given uncertainties in these parameters.

\subsection*{Optical Tweezer trapping lifetimes}

\begin{figure}[t!]
	\centering
	\includegraphics[scale=1]{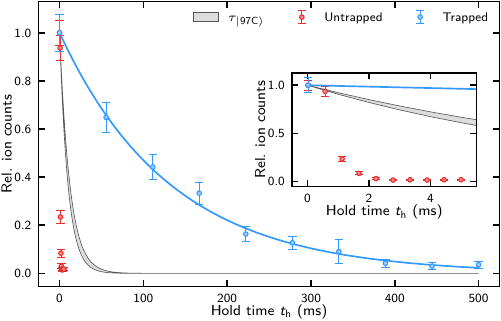}
	\caption{Rydberg atom tweezer lifetime. Measured loss of all detected Rydberg states (total ion signal on the MCP) as a function of the hold time $t_\mathrm{h}$ for initial preparation of $\ket{97\mathrm{C}}$ with the tweezer kept on (blue circles) and switched off (red circles). Ion counts are normalized to the initial value. The solid line is an exponential fit to the data to extract the 1/e trapping lifetime. Turning the tweezer off at $t_\mathrm{h}=0$ results in rapid loss of the signal within a millisecond (see inset), because the Rydberg atoms leave the trapping region and are no longer detected by our SSFI protocol. Errorbars indicate one standard error of the mean.}
	\label{fig:tweezer_lifetimes}
\end{figure}

The measurement of these extreme lifetimes requires that we keep the Rydberg atoms trapped and that atom loss from the tweezer is negligible. 
Trapping is also essential when exploiting the long lifetimes for quantum simulation or computing. 
Typically, Rydberg atoms are expelled from standard Gaussian optical tweezers because of the repulsive ponderomotive force on the Rydberg electron \cite{Barredo2020,Ravon2023}. 
Here, we use alkaline-earth atoms, where the polarizable Sr$^+$ ionic core provides a sufficiently strong attractive potential to confine our CRS in standard Gaussian tweezer~\cite{Wilson2022,Hoelzl2024}.
Moreover, trapping does not require a circular Rydberg orbit but remains effective also when the Rydberg electron slowly decays into other Rydberg levels with different angular momentum or principal quantum number. 
To characterize the trapping lifetime, we initialize the atom in $\ket{97\mathrm{C}}$ as before, but now measure the total number of detected ions as a function of $t_\mathrm{h}$ without differentiating between product Rydberg states in different $n$ manifolds (Fig.~\ref{fig:tweezer_lifetimes}). 
An exponential fit of the data yields a 1/e trap lifetime of \SI{133(6)}{\milli\second}, about an order of magnitude longer than the initial state lifetime. 
A control measurement with the tweezer trap turned off (inset of Fig.~\ref{fig:tweezer_lifetimes}) shows that without trapping, the ion signal fades after a millisecond. 
Notably, this measurement gives a lower bound for the trapping time, because of the limited detection range of Rydberg levels arriving at the MCP for the chosen SSFI ramp in combination with ion-optical constraints. 
For example, with the SSFI parameters used here, Rydberg states with $n>104$ are not detected. 
Similarly, decay to significantly lower angular momentum states also leads to ionization thresholds, that are outside of the SSFI detection range. 

\section*{Discussion}
To conclude, we have observed Purcell-enhanced lifetimes at record high-$n$ trapped CRS of more than \SI{10}{\milli\second} in a room-temperature optical-tweezer apparatus. These are the longest-lived Rydberg states produced in a laboratory so far, even exceeding recent achievements in a cryogenic environment \cite{CantatMoltrecht2020}. We characterized the Purcell effect on blackbody radiation far beyond the cut-off condition of our suppression structure via the enhancement of circular state lifetimes over a wide range of coherently connected $n$. This revealed the effects of additional resonant modes inside our electric field control in agreement with our theoretical models, demonstrating important aspects in the design of blackbody-radiation suppressing structures.

Combined with the recently demonstrated control of interactions for CRS \cite{Mehaignerie2025}, our results pave the way for realizing a novel Rydberg-based quantum simulator capable of boosting simulation times by orders of magnitude, augmented by local optical addressing via the optical active ionic core \cite{Muni2022,Wirth2024}. 
The next steps to fully exploit the long lifetimes for quantum simulations require to also push the $T_2$ coherence time for qubits encoded in circular states into the millisecond range. In our setup technical sources, specifically magnetic field noise and electric field homogeneity, currently limit this $T_2$ time to the \SI{100}{\micro\second} scale ~\cite{Hoelzl2024}, which can be improved by active feedback and careful cancellation of gradients. Thermal trap-induced dephasing can also be addressed exploiting laser cooling on the ionic core~\cite{Bouillon2024}. Furthermore, dynamical decoupling provides additional intriguing perspectives~\cite{Holland2023, Choi2020}, particularly regarding dephasing from static gradients and the comparatively slow thermal motion in the tweezers. Overall, we expect that with these measures coherence times on the scale of the circular state lifetime are within reach.

The demonstrated control over a wide range of coherently connected CRS opens perspectives for implementing circular electron qudits or synthetic dimensions \cite{Kanungo2022}. 
We have not yet explored the even larger state space available when including also elliptical states, which has recently been employed for sensing applications \cite{Ramos2017}, and proposed for encoding large-spin systems \cite{Kruckenhauser2023}, or effective dual-species models \cite{Dobrzyniecki2025}. The hundreds of millisecond trapping times in combination with the alkaline-earth atom explored here, opens up exciting possibilities for coherent motional control of individual or interacting Rydberg atoms \cite{Bouillon2024,Mehaignerie2023}, or high-precision ion clock spectroscopy inside a Rydberg orbit \cite{Ludlow2015}.

\section*{Methods}
\renewcommand{\thesection}{\Alph{section}}

\section{Circular State Control}\label{sec:spectroscopy}

The resonances $n\rightarrow n'=n+2$ as shown in Fig.~\ref{fig:Fig2} were determined by scans of the frequency $f$ of the last microwave (MW) pulse. The applied electric field is calibrated using the $\ket{79\mathrm{C}} \leftrightarrow \ket{81\mathrm{C}}$ transition.
Throughout all measurements a magnetic field of \SI{1.5(1)}{\gauss} was applied, which is also considered in the theory curves in Fig.~\ref{fig:Fig2} by an additional Zeeman shift.
The MW drive is emitted from a horn antenna with a linear polarization before entering the electric field control.
At the position of the atoms the MW field has an uncontrolled but stable polarization.
he populations $p_n$ and $p_{n'}$ of the initial and final state, respectively, were extracted from the arrival time windows of the ions after SSFI (see Fig.~\ref{fig:Fig1}(d)).
We perform a fit to the population contrast $p_{n'} - p_n$ to obtain the resonance frequencies $f_0$ and full width at half maximum (FWHM).

Up to $n=95$ the spectroscopy was typically performed with near $\pi$-pulses at fixed electric fields $E_z$.
In these cases a Gaussian was fitted to the data.
However, due to the increased suppression of our MW drive inside the electrode structure below the cut-off frequency, $f_\mathrm{cut}=c/(2d)$, the desired $\sigma^+\sigma^+$ transitions became narrower and particularly challenging to locate for larger $n$.
In these cases, an electric field ramp of amplitude $\Delta E_i$ symmetric around the set field $E_i$ was performed during the pulse to drive Landau-Zener sweeps instead.
The resulting population contrast then appears as a smoothened rectangle function and was modeled as

\begin{align}
	\frac{A}{2} \left[ \erf\left(\frac{f - f_0 + w/2}{\sigma}\right) + \erf\left(\frac{f_0 - f + w/2}{\sigma}\right) -1\right] + b
\end{align}

in order to fit the resonance, with amplitude $A$, FWHM $w$, standard deviation $\sigma$ of the errorfunction $\erf$, and offset $b$.

\begin{figure}[tb!]
\centering
	\includegraphics[scale=1]{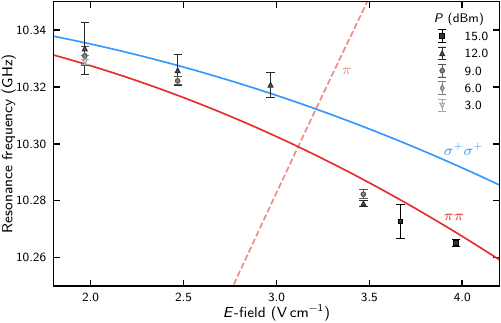}
	\caption{Single-photon lightshifts. Measured resonances of two-photon transitions from $\ket{85\mathrm{C}}$ to $n=87$ compared to the transition $\sigma^+\sigma^+$ resonance between CRS (blue), and the nearby $\pi\pi$ transition to an elliptical state (red), following the same procedure as in Fig.~\ref{fig:Fig2}. Different shades and markers indicate different power settings of the MW source. The data points are lightshifted away from the $\pi\pi$ resonance (red line) by a single-photon $\pi$ transition, connecting the excited state to $n=88$ (red dashed line). Errorbars denote the FWHM of the measured resonances.}
	\label{fig:lightshifts}
\end{figure}

For each $\mathrm{LZ}_i$ pulse the ramp and pulse parameters were mainly optimized for fast population transfer.
All determined parameters for the CRS preparation are summarized in Table~\ref{tab:pulse_parameters}.
With these settings the fidelity for the individual Landau-Zener steps, defined as the fraction of transferred population from $n$ to $n'$, reaches \SI{94}{\percent} on average. We believe that by more sophisticated parametrization of the voltage ramps and MW pulses or increased microwave power, the transfer can be further optimized. The finite LZ transfer leads to residual populations remaining in CRS of intermediate $n$ throughout the ladder, which is considered in the lifetime evaluations by taking all initial populations in the detected range into account (see Methods \ref{sec:SSFI}).
The use of a vector signal generator as MW source ensures ultrafast switching times of  $\SI{20}{\micro\second}$ between pulses of different frequencies and power.
Thereby the coherent population transfer up to $n=101$ takes approximately \SI{400}{\micro\second}.
Notably, this is already longer than the free-space lifetime of the lowest used CRS.

\begin{table*}[!tb]
\centering
	\caption{State preparation parameters. Parameters used for the preparation of CRS up to $n=103$ via subsequent Landau-Zener sweeps $\mathrm{LZ}_{i}$. This includes the set frequency $f_i$, pulse duration $T_i$ and power $P_i$ of the MW pulses as well as voltages $U_i$ applied between the ITO electrodes. During each pulse the voltage is ramped by $\Delta U_i$ symmetrically around $U_i$ from larger to lower values.}
	\label{tab:pulse_parameters}
		\begin{tabular}{lllllll}
			\toprule
			$i$ & Transition & $f_i(\si{\giga\hertz})$ &$U_i (\si{\volt}) $ & $\Delta U_i (\si{\milli\volt}) $ & $T_i (\si{\micro\second})$ & $P_i (\si{\deci\bel\of{m}})$ \\
			\midrule
			0& $\ket{79\mathrm{C}} \rightarrow \ket{81\mathrm{C}}$ & 12.83487 & 3    & 220 & 12 & 0  \\
			1& $\ket{81\mathrm{C}} \rightarrow \ket{83\mathrm{C}}$ & 11.92764 & 2    & 300 & 12 & 9  \\
			2& $\ket{83\mathrm{C}} \rightarrow \ket{85\mathrm{C}}$ & 11.094	  & 2    & 400 & 8  & 9  \\
			3& $\ket{85\mathrm{C}} \rightarrow \ket{87\mathrm{C}}$ & 10.23831 & 5.53 & 220 & 13 & 12  \\
			4& $\ket{87\mathrm{C}} \rightarrow \ket{89\mathrm{C}}$ & 9.525315 & 5.69 & 30  & 40 & 15  \\
			5& $\ket{89\mathrm{C}} \rightarrow \ket{91\mathrm{C}}$ & 8.891    & 5.5  & 150 & 10 & 15 \\
			6& $\ket{91\mathrm{C}} \rightarrow \ket{93\mathrm{C}}$ & 8.36933  & 4    & 150 & 10 & 15 \\
			7& $\ket{93\mathrm{C}} \rightarrow \ket{95\mathrm{C}}$ & 7.83447  & 4    & 100 & 15 & 15 \\
			8& $\ket{95\mathrm{C}} \rightarrow \ket{97\mathrm{C}}$ & 7.34     & 4    & 400 & 10 & 15 \\
			9& $\ket{97\mathrm{C}} \rightarrow \ket{99\mathrm{C}}$ & 6.909    & 3.5  & 160 & 30 & 15 \\
			10& $\ket{99\mathrm{C}} \rightarrow \ket{101\mathrm{C}}$ & 6.4888 & 3.5  & 150 & 30 & 15 \\
			11\footnotemark[1]& $\ket{101\mathrm{C}} \rightarrow \ket{103\mathrm{C}}$ & 6.0975 & 3.5  & 150 & 150 & 15 \\
			\bottomrule
		\end{tabular}
		\flushright\footnotetext[1]{This transition was only probed spectroscopically with the provided settings and not optimized.}
\end{table*}

Throughout this work, two-photon transitions are used for state preparation.
This can make the excitation scheme sensitive to disturbing single-photon transitions crossing the two-photon resonance.
Such an example is shown in Fig.~\ref{fig:lightshifts} for the transition $\ket{85\mathrm{C}}$ to $n=87$, measured without LZ sweeps.
Here, a linearly Stark shifted single-photon $\pi$ excitation to an elliptical state of the next higher manifold $n=88$ (dashed line) crosses the two-photon transitions.
This results in a up to tens of \si{\mega\hertz} shift of the measured resonance around this crossing.
At field values between 2 and \SI{3}{\volt\per\centi\meter}, the feature appears around the $\sigma^+\sigma^+$ transition to the target CRS (blue line),
but approaches the $\pi\pi$ transition (red line) when reducing the MW power.
This reveals that the measured resonance originates from the $\pi\pi$ transition and makes it challenging to unambiguously identify the $\ket{n\mathrm{C}}$ to $\ket{n+2\mathrm{C}}$ resonance.
More transitions are expected to cross the two-photon resonance at lower electric fields.
To keep the Landau-Zener ladder unperturbed, we thus select field values avoiding such scenarios, specifically for the range of $n$, where our electrode structure enhances $\pi$ polarization (\textit{cf.} Fig~\ref{fig:Fig3}(e)).

\begin{figure*}[tb!]
	\centering
	\includegraphics[scale=1]{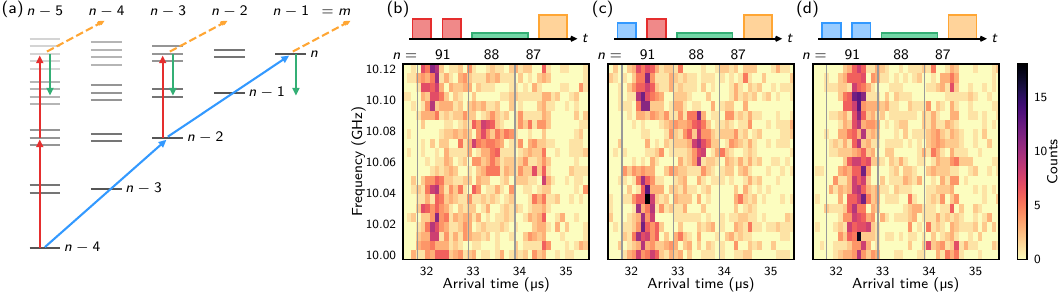}
	\caption{Circular state validation scheme. (a) Level scheme illustrating the relevant transitions used to validate if a prepared state with principal quantum number $n$ is indeed the desired CRS $\ket{n\mathrm{C}}$. (b)--(d) Experimental demonstration of the validation scheme at the example of $\ket{89\mathrm{C}}$ preparation. The top row of each panel depicts the applied pulse sequence, starting from $\ket{85\mathrm{C}}$. Colors of the indicated pulses correspond to the transitions shown in (a). The first two pulses prepare elliptical states for (b) and (c), and the desired CRS for (d). The datasets below show frequency scans of a subsequent single-photon Landau-Zener probe pulse (green) to a linearly Stark shifted elliptical state with $n = 88$. The presence of a resonance in (b) and (c) together with the absence of a resonance in (d) is the signature for successful preparation of $\ket{89\mathrm{C}}$. A final shelving pulse to $n = 91$ (orange) is applied to increase the arrival time contrast.}
	\label{fig:CRS_check}
\end{figure*}

In addition, we devised an independent check that the Landau-Zener sweeps end up in the desired CRS.
This is illustrated in Fig.~\ref{fig:CRS_check} for the preparation of $\ket{89\mathrm{C}}$.
In principle a CRS can be unambiguously identified by the absence of a linearly Stark shifted $\pi$ transition (green) to lower $n$.
To exclude technical reasons of an absent resonance, we take reference measurements by intentionally preparing an elliptical state in $n$ via a two-photon $\pi\pi$ excitation (red).
This is done for multiple initial states (here, $m=n-3$ and $m=n-5$), which feature the same type of $\pi$ transition (green) with comparable dipole matrix elements.
Importantly, those resonances have the same linear Stark shift, and are thus expected at very similar frequencies.
A detection of both lines via identical Landau-Zener sweeps (Fig.~\ref{fig:CRS_check}(b) and (c)) together with no detected resonance when preparing the assumed CRS (Fig.~\ref{fig:CRS_check}(d)), unequivocally confirms successful state preparation.
Note that an additional shelving pulse to a higher principal quantum number (here, another $\sigma^+ \sigma^+$ transition to $n+2$) is added (dashed yellow) for larger separation in arrival time after SSFI.

\section{State selective field ionization}\label{sec:SSFI}

For state selective field ionization (SSFI), we apply a linear high-voltage ramp to two of the ring-shaped electrodes (to the right in Fig.~\ref{fig:Fig1}(a)) which ionizes the Rydberg atom and accelerates the ion toward the MCP detector.
The MCP is located on one side of the electrode setup (outside the visible region in Fig.~\ref{fig:Fig1}(a) toward positive $z$).
Ion trajectories are manipulated by additional fields between the ITO plates and further deflector electrodes in the detection region.
The SSFI scheme has been optimized, in order to enable the detection of a large range of $n$ required in this work.
This involved adiabatic rotation of the quantization from $z$ to $x$ axis prior to ionization, as indicated in Fig.~\ref{fig:Fig1}(b), and adjustments in deflector settings.
Two separate sets of SSFI parameters were determined for the measurement of the CRS lifetimes (Fig.~\ref{fig:Fig3}).
One set allowed for simultaneous detection of $n$ between 79 and 93 and was used for measurements up to $\ket{89\mathrm{C}}$.
For all higher-lying CRS the second SSFI parameter set for detection between $n=89$ and $n=104$ has been used).
The populations of the respective states are determined from the detected ion counts in all accessible bins with the used ionization ramp. This includes residual elliptical states in the $n=79$ bin, which are not addressed by our MW pulses (see also Fig.~\ref{fig:Fig1}(d)). Consequently, the initial target circular state populations of $n > 79$ for the lower CRS branch (Fig.~\ref{fig:Fig3}(c)) do include the \SI{70}{\percent} fidelity of CRS preparation~\cite{Hoelzl2024}, while the higher CRS branch (Fig.~\ref{fig:Fig3}(d)) does not.

While our SSFI scheme mainly distinguishes principal quantum numbers $n$, we note that the intra-manifold states ionize slightly differently.
Generally, states with lower angular momentum tend to ionize at lower fields because of a reduced centrifugal barrier.
In addition, the linear Stark effect can further affect the ionization threshold \cite{Gallagher1994}.
Indications of this are, for example, visible in Fig.~\ref{fig:CRS_check}(b)--(d) in the $n=91$ window. 
Here, the ion signal stemming from elliptical Rydberg states (Figs.~\ref{fig:CRS_check}(b) and (c)) is shifted to lower arrival times (on the order of \SI{100}{\nano\second}) when compared to the signal from the CRS (Fig.~\ref{fig:CRS_check}(d)).
A second example is apparent in the integrated ion signal on top of Fig.~\ref{fig:Fig3}(a). 
Enhanced $\pi$ polarized decay for this $n$ favors population of elliptical states, which can lead to a substructure within the $n$-manifold windows. In order to unambiguously assign arrival time windows for all $n$, we take histograms of all measurements with the same SSFI settings into account.

\section{Purcell Suppression Factors}\label{sec:purcell}

\begin{figure}[tb!]
	\centering
	\includegraphics[scale=1]{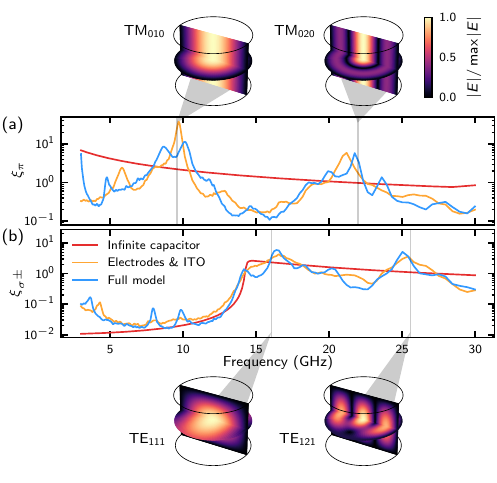}
	\caption{Purcell factors of the BBR suppression capacitor. Calculated Purcell factors $\xi$ for $\pi$ (a) and $\sigma^\pm$ (b) polarized modes. The analytical result for an infinite ITO capacitor (red) with a spacing of $d=\SI{10.5}{\milli\meter}$ reveals the suppression of $\sigma^\pm$ modes below $f_\text{cut}$. Including the ring-shaped electrodes with an inner radius of $a=\SI{12}{\milli\meter}$ in our simulation (orange) results in additional resonances, most prominent in (a). Those can be assigned to modes of a cylindrical cavity, with mode profiles depicted in the insets (dimensions not to scale). The corresponding resonance frequencies are marked by vertical gray lines. The full model (blue) includes additional conductive elements which further affect the resonances.}
	\label{fig:cavity_models}
\end{figure}

The Purcell suppression factors for our capacitor structure are simulated with COMSOL Multiphysics using the RF module.
A small perfectly conducting electric dipole antenna is placed at the atom position. 
For a small dipole antenna, the Purcell factor $\xi$ can be calculated as \cite{Krasnok2015}
\begin{equation}
	\xi = \frac{\operatorname{Re}(Z)}{\operatorname{Re}(Z_0)} \, ,
\end{equation}
where $Z$ is the simulated impedance with the suppression capacitor structure placed around the dipole and $Z_0$ is the impedance in free space.
To circumvent finite size effects, a perfectly matched boundary layer is placed around the simulated structure.
By placing the small dipole antenna parallel (perpendicular) to the ITO plates, the Purcell factor for microwaves with $\sigma^\pm$ ($\pi$) polarization in the atomic frame are obtained.
For the full model simulations, the four stainless steel ring electrodes, the ITO cavity and the stainless steel support structure (see Fig.~\ref{fig:Fig1}(a)) are taken into account.
We assume the stainless steel parts as perfect conductors and extract the conductivity $\sigma$ of the ITO coatings from the layer thickness $l_\mathrm{ITO} = \qty{700}{nm}$ and the sheet resistance $R_\square = \SI{2.4}{\OhmSq}$, where $\si{\OhmSq}$ denotes the unit ohms per square. 
This was measured before the ITO plates were put into vacuum and results in $\sigma = 1/(l_\mathrm{ITO} R_\square) = \SI{0.595}{\mega\siemens\per\meter}$.
These parameters result in a microwave reflectivity of the ITO plates of about \SI{96}{\percent}, extrapolated from measurements on comparable samples \cite{Meinert2020}.

The results of these simulations are shown in Fig.~\ref{fig:cavity_models}.
In comparison to analytical calculations for an infinitely large ITO capacitor (red) \cite{Meinert2020}, the retrieved Purcell factors $\xi$ for the fully modeled electrode structure (blue) show additional resonances. 
Those especially affect $\xi_\pi$ below the cutoff frequency for the plane capacitor $f_\text{cut} = c/(2d)$, where $c$ denotes the speed of light.
Here, $\pi$ polarized fields experience a significant enhancement, which is attributed to the formation of $\mathrm{TM}_{ijk}$ modes of a cylindrical cavity~\cite{Pozar2012}, formed by the ITO sheets and the ring electrodes with an inner radius of $\SI{12}{\milli\meter}$. 
The indices $i$, $j$ and $k$ represent the azimuthal, radial and longitudinal mode numbers.
Restricting the simulation to said components (orange) reveals that the observable enhancement features closely match the analytical solutions of both TE and TM modes, marked by gray vertical lines.
The corresponding mode structures, as illustrated in Fig.~\ref{fig:cavity_models}, all have intensity maxima in the center of the cylindrical cavity.
Specifically, the $\mathrm{TM}_{010}$ and $\mathrm{TM}_{020}$ modes appear in $\xi_\pi$, while $\mathrm{TE}_{111}$ and $\mathrm{TE}_{121}$ result in additional peaks in $\xi_{\sigma^\pm}$ above $f_\mathrm{cut}$.
Additional smaller features are due to remaining conductive components of the support structure, as well as the specific shape of the ring electrodes.

\section{Circular State Lifetime Calculations}\label{sec:calculations}

The analytical lifetime $\tau$ of a CRS $\ket{n\mathrm{C}}$ is obtained from the decay rate, $\Gamma = 1/\tau$, which is given by

\begin{equation}\label{eq:Decay}
	\Gamma = \sum_{\ket{s}} \Gamma_{\ket{s}} = \sum_{\ket{s}} \left( \bar{n}(\omega,T) + \delta_{\ket{s},\ket{n-1\mathrm{C}}} \right) \xi A_{\ket{n\mathrm{C}}, \ket{s}} \, .
\end{equation} 

Here, $A_{\ket{n\mathrm{C}}, \ket{s}}$ denotes the Einstein $A$ coefficient between states $\ket{n\mathrm{C}}$ and $\ket{s}$, $\omega$ the corresponding transition frequency, and $\bar{n}$ the thermal average photon number at temperature $T$. The sum runs over all final states $\ket{s}$ with non-zero dipole matrix element entering $A$, and the Kronecker symbol  $\delta_{\ket{s},\ket{n-1\mathrm{C}}}$ accounts for the only spontaneous decay channel, $\ket{n\mathrm{C}} \rightarrow \ket{n-1\mathrm{C}}$. The effect of the electrode structure enters via the Purcell factor $\xi$ (see Methods~\ref{sec:purcell}, $\xi=1$ for decay in free space).

Our lifetime measurements are conducted in the presence of the electric field $E_\text{h}$. We thus calculate the lifetime also in a finite electric field. To this end, we use analytic expressions for hydrogen, which are accurate for the high orbital angular momentum states explored here. 
First, the electric field affects the transition frequencies $\omega$ via the Stark shifts, which we calculate up to fifth order in $E_\text{h}$ \cite{Stebbings2011}. 
Note that this expression is also used for the theory curves in Fig.~\ref{fig:Fig2}.
Second, the dipole matrix elements entering $A$ need to be calculated between energy eigenstates in the electric field. For this, it is convenient to start from the parabolic basis states $\ket{nkm}$, i.e. solutions for the electron wavefunction in the electric field up to first order in $E_\text{h}$, with a dipole moment proportional to the quantum number $k$. The dipole matrix element $d_{nkm}^{n'k'm'}$ between parabolic states $\ket{nkm}$ and $\ket{n'k'm'}$ is then related to the dipole matrix elements $d_{nlm}^{n'l'm'}$ between zero field hydrogen orbitals $\ket{nlm}$ and $\ket{n'l'm'}$ via
\begin{align}
	d_{nkm}^{n'k'm'} =  \sum_{l=m}^{n-1} \braket{nkm|nlm} \sum_{l'=m'}^{n'-1} \braket{n'k'm'|n'l'm'}d_{nlm}^{n'l'm'} \, ,
\end{align}
where $\braket{nkm|nlm}$ denote the Clebsch-Gordan coefficients to describe elliptical states \cite{Gallagher1994}. The values for $d_{nlm}^{n'l'm'}$ can again be derived analytically in the hydrogen case. Specifically the radial matrix elements starting from a CRS can be evaluated with the Gordon formula \cite{Dewangan2012}. Finally, both the transition frequencies and dipole matrix elements are scaled according to the Rydberg constant $\mathrm{Ry}_{^{88}\mathrm{Sr}} = \SI{109736.631}{\per\centi\meter}$ \cite{Couturier2019}, to account for the nuclear mass of the $^{88}\mathrm{Sr}$ atom.

When computing $\Gamma$, it is sufficient to cap the sum over states $\ket{s} = \ket{n'k'm'}$ in Eq.~\eqref{eq:Decay} at $\Delta n_\mathrm{max} = |n'-n| = 3$, because of the rapid decrease of the relevant dipole matrix elements with $\Delta n$.

\section{Validation of the Rate Model}\label{sec:rate_model}
\begin{figure}[tb!]
	\centering
	\includegraphics[scale=1]{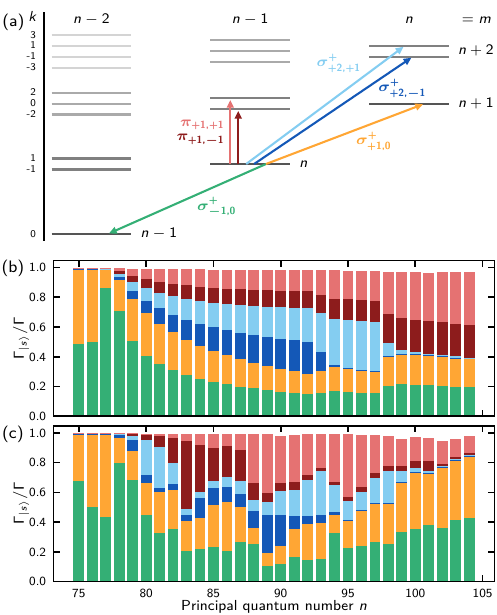}
	\caption{Circular state decay contributions in the suppression capacitor. (a) Level scheme with relevant decay channels of a CRS $\ket{n\mathrm{C}}$, labeled as $(\pi/\sigma^+)_{\Delta n, \Delta k}$ where $\Delta n$ and $\Delta k$ denote the change of principal and parabolic quantum numbers respectively. (b, c) Calculated relative decay rates $\Gamma_{\ket{s}}/\Gamma$, including Purcell factors for the infinite ITO capacitor (b) and the fully modeled electrode structure (c). The bar colors correspond to the arrow colors indicating the transitions in (a). The calculations assume an applied electric field $E_\text{h}=\SI{5.9}{\volt\per\centi\meter}$, lifting the degeneracy of elliptical states with different $k$.}
	\label{fig:decay}
\end{figure}

As stated in the main text, the CRS lifetimes are obtained by fitting a rate model to the experimental data (see Fig.~\ref{fig:Fig3}).
This model assumes a ladder of states, each corresponding to a principal quantum number $n$, which are connected by a transition rate to neighboring states $n' = n\pm 1$.
In free space this method works trivially well \cite{CantatMoltrecht2020}, because the dipole matrix element for $\sigma^+$ transitions to neighboring CRS by far dominates over all other decay paths, effectively limiting the experimentally explored state space to a single ladder of CRS.
The BBR mode density inside the electrode structure (Fig.~\ref{fig:cavity_models}), however, modifies the decay as illustrated in Fig.~\ref{fig:decay}(b).
On the one hand, the two enhanced $\pi$ transitions to elliptical states of the next higher manifold ($\Delta n =1$, red) can reach comparable and even stronger couplings than the suppressed $\sigma^+$ channels ($\ket{n\mathrm{C}} \rightarrow \ket{n\pm1\mathrm{C}}$, green/orange). 
On the other hand, for sufficiently low $n$, two $\sigma^+$ transitions to elliptical states with $\Delta n = +2$ (blue) are not yet suppressed because of their wavelength, which is about half as short, and can also contribute significantly to the total decay rate.

All of these effects are summarized in Fig.~\ref{fig:decay}(b) and (c), showing the relative contributions of the modified decay rates $\Gamma_{\ket{s}}$ to the total decay rate $\Gamma$ as a function of $n$.
For an infinite ITO capacitor (Fig.~\ref{fig:decay}(b)), one finds a steady change in $\Gamma_{\ket{s}}/\Gamma$, starting at $n=77$ until the two above-mentioned $\Delta n = +2$ transitions are suppressed at around $n=94$ and $n=99$.
Note that suppression for the two transitions sets in at different $n$ because of the different Stark shifts of the elliptical states in the applied electric 
field $E_\text{h}=\SI{5.9}{\volt\per\centi\meter}$.
For the majority of $n$ values, we see that the overall decay significantly populates non-circular states, which increases the relevant state space in each hydrogenic manifold.
In the case of the full electrode structure (Fig.~\ref{fig:decay}(c)), the additional resonances in $\xi$ result in a highly $n$-dependent behavior of $\Gamma_{\ket{s}}/\Gamma$.
For example, in the extreme case of $n=89$, around \SI{80}{\percent} of the CRS decay products are in elliptical states, while for $n>100$ the same fraction of the population remains in the ladder of CRS. Once in an elliptical state, the decay continues with different rates than for CRS with branching to even more substates of the hydrogenic manifolds.

\begin{figure}[b!]
	\centering
	\includegraphics[scale=1]{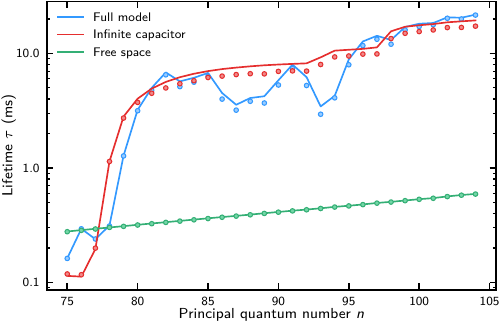}
	\caption{Rate model characterization. The calculated lifetime of CRS at different principal quantum numbers (lines) is compared to the lifetime extracted from a fit of our rate model to numerically simulated decay dynamics (dots) for different BBR environments as indicated.}
	\label{fig:ratemodels}
\end{figure}

This analysis raises the question, to what extent the simplified rate model introduced above allows for accurately determining the CRS lifetimes. One may be tempted to extend the model to more states, but we find that this leads to overfitting of the experimental data. Instead, we show in the following that our approach allows for reliable extraction of lifetimes. To this end, we start with numerically simulating the CRS decay dynamics in its full complexity. This is done by solving an extended rate equation model, which includes additional 229 elliptical states besides the CRS ladder. Transition rates are computed from the analytic hydrogen model as introduced in Methods~\ref{sec:calculations}. Note that rates between states with $l<n-1$ can be calculated via additional recursion formulas \cite{Dewangan2012}. From the numerical results, we sum up all state populations belonging to the same $n$-manifold, which effectively mimics the experimental data. Such simulated decay dynamics for $n$-manifolds are then fitted with our simplified rate model. The CRS lifetimes extracted from the fits are shown in Fig.~\ref{fig:ratemodels} for different BBR environments and compared to directly calculated lifetimes of $\ket{n\mathrm{C}}$ using Eq.~\ref{eq:Decay}. As expected, for ranges of $n$ dominated by $\sigma^+$ decay to neighboring CRS (\textit{cf.} Fig.~\ref{fig:decay}), the fit results (dots) are very close to the calculated lifetime (lines). Importantly, we find only small deviations ($<\SI{16}{\percent}$) over the entire range of $n$ values, even for the cases with significant elliptical state population. Additionally, the fit yields systematically lower values for the lifetimes. These findings support and quantitatively validate our experimental data analysis.

\section*{Data Availability}
The ion arrival time data generated in this study are openly available at Zenodo (\url{https://doi.org/10.5281/zenodo.20842521})\nocite{Pultinevicius2026}

\section*{Acknowledgements}
We thank Tilman Pfau for support over the past years. 

\section*{Funding}
We acknowledge funding from the Federal Ministry of Research, Technology and Space under the Grants CiRQus and QRydDemo, and the Horizon Europe Programme HORIZON-CL4-2021-DIGITAL-EMERGING-01-30 via Project No. 101070144 (EuRyQa).

\section*{Author Contributions}
E.P., A.G., F.T., and C.H. designed and carried out the experiments. E.P. and C.H. performed the data analysis and simulations. E.P. and F.M. wrote the manuscript with contributions from all authors. F.M. supervised the project.

\section*{Competing interests}
The authors declare no competing interests.

\end{document}